\documentstyle[11pt,paspconf,epsf]{article}

\begin{document}

\title{ The Massive Nearby SBNGs: Young Galaxies in Formation}
\author{Roger Coziol}
\affil{ Laborat\'orio Nacional de Astrof\'\i sica, Itajub\'a, Brasil}
\author{Thierry Contini}
\affil{School of Physics \& Astronomy, Tel-Aviv University, Israel}
\author{Emmanuel Davoust}
\affil{ UMR 5572, Observatoire Midi-Pyr\'en\'ees, Toulouse, France}
\author{Suzanne Consid\`ere}
\affil{UPRES-A 6091, Observatoire de Besan\c con, France}

\begin{abstract}

We discuss the implications of the recent discovery that Starburst 
Nucleus Galaxies (SBNGs) have lower oxygen abundances than ``normal galaxies" 
of the same morphological type.  Our interpretation of this result is that 
SBNGs are young galaxies, still in the process of formation.  
We consider several alternatives, 
but none of them provides a viable explanation.
This new result has important consequences for our understanding of
galaxy evolution, as it confirms the scenario of hierarchical formation of
galaxies, and explains recent observations of the {\it Hubble Space
Telescope}.  

\end{abstract}

% KEYWORDS SHOULD BE INCLUDED, BUT THEY ARE NOT PRINTED IN THE HARDCOPY!

\section{Introduction}

The current paradigm about starbursts is that they
are sporadic events that can occur at any time in the evolution of a
galaxy, generally triggered by galaxy interactions (Heckman 1997). 
One of the reasons for this assumption is probably that 
starbursts are often seen in gravitationally interacting galaxies,
and this paradigm has been reinforced in the last decade by the
discovery that ultraluminous infrared galaxies are generally violently
interacting.  

In a recent paper (Coziol et al.\,1997a), we gathered published and
unpublished data on several large samples of Starburst Nucleus Galaxies
(SBNGs) to show that they have lower oxygen abundance than normal
galaxies of the same morphological type, and that early-type SBNGs are
even more deficient in oxygen than late-type ones.

The simplest interpretation of this result is that SBNGs are young
galaxies, which are still in the process of formation.  The fact that
early-type galaxies are less abundant in oxygen than late-type ones can
be readily understood in the frame of the theory of hierarchical
formation of galaxies.  In this scenario of galaxy evolution (Tinsley \&
Larson 1979; Struck-Marcell 1981), ellipticals and bulges of
spirals are formed by mergers of stellar and gaseous systems, while
disks form by the collapse of the gas left from the successive mergers.

\section{Alternative interpretations of the low oxygen abundances}

Is our interpretation of SBNGs as young galaxies inescapable, or are
there viable alternative explanations for the low values of the
oxygen abundances measured in SBNGs?  We review here four possible
hypotheses.

$\bullet$ The first alternative that comes to mind for explaining that 
a galaxy has an unusually low oxygen abundance is rejuvenation by
accretion of unprocessed gas during gravitational interaction with 
another galaxy. But most of the galaxies in our samples are isolated 
(Coziol et al.\,1995; Contini 1996), and so are most HII galaxies 
(Telles \& Terlevich 1995). 
  
We have examined several samples of interacting galaxies 
to see if there is a relation between interaction and oxygen abundance.
If one removes all the Markarian galaxies
from the sample of interacting galaxies of Keel et al.\,(1985), 
the remaining galaxies have solar abundances on the average. 
Therefore, interactions do not necessarily imply low metallicities.   
The interacting galaxies of Bushouse (1986) do have low abundances, 
but these galaxies are also IRAS sources.  They are thus probably of 
the SBNG type, because $\sim$ 85\% of the SBNGs are IRAS sources 
(Coziol et al. 1995). This suggests that we should distinguish 
two types of interacting galaxies at the present epoch, metal poor
ones, which are young galaxies still in formation ($\sim1/4$ of the SBNGs 
and HII galaxies show signs of interaction), and metal rich ones, 
which are old and well evolved galaxies. 

In fact, we should expect to find few interacting galaxies among the SBNGs,
if the latter are the outcome of multiple mergers of small mass
galaxies that took place 2 or 3 Gyr ago. Obviously, the remnant of a merger 
must be a relatively massive and isolated galaxy. 

The unprocessed gas could also come from intergalactic space or a massive halo.
The presence of a bar is then required to provide the gravitational torque 
for funneling the gas toward the center.
But this would only work for part of the sample, since 
not all our galaxies are barred. And this mechanism leads to a paradox,
since it seems to be less
efficient in late-type galaxies, where bars are more frequent.
One would also have to explain why gas is accreted in some galaxies (the
SBNGs) and not in others (the normal ones).

$\bullet$ Another possibility is winds from massives stars and 
supernova explosions;
they could have swept away the enriched gas formed in the nuclear
regions of SGNGs.  
One then expects the radial oxygen abundance gradients in SBNGs to be
flat, or very low. But recent high signal-to-noise spectroscopic 
observations of 24 galaxies from the sample of Contini et al.\,(1998)
reveal that the abundance gradients in barred SBNGs are not unusual
(Consid\`ere et al., in preparation).
Furthermore, De Young \& Heckman (1994) find that gaseous outflows tend to
destroy low mass galaxies, but probably play no role in the evolution of
massive ones. Also, why would stellar winds be stronger 
in early-type galaxies (which show lower abundances) than in late-type ones?

$\bullet$ One could also try to explain our results by advocating a
different source of ionization for the gas in the nuclear regions.
Our estimates of the oxygen abundances rest on the assumption that the
observed emission lines are produced by OB stars which excite the
interstellar gas. If this assumption is incorrect, for example if the source of
ionization of the gas is nonthermal, such as a hidden AGN or
shock heating, then of course the lower abundances are not
real.
The only peculiar characteristic of our SBNGs is a high 
[N\,{\sc ii}]\,$\lambda$6584/H$\alpha$ ratio (see also Coziol et al.\,1997b), 
which could indeed point to non-thermal processes.
We have shown elsewhere (Coziol et al.\,1997a; Contini et al.\,1998)
that this excess emission of nitrogen in our sample 
is not the sign of AGN activity.
We believe that the relative excess of [N\,{\sc ii}] corresponds to 
an overabundance of nitrogen which is produced by the evolution of
intermediate mass stars formed during a sequence of bursts 
over the last 2-3 Gyr (Carlos Reyes \& Coziol, in preparation).

$\bullet$ Finally, our abundance estimates could also be incorrect if dust 
removes atomic coolants from the gas phase.
Calculations by Shields \& Kennicutt (1995) indicate
that dust can have an appreciable influence on the optical spectrum in
environments with high (above solar) metallicity. The predicted line
ratios of [N\,{\sc ii}]/H$\alpha$ \ agree \ reasonably \ well 
\ with \ our \ high-abundance \ nuclei
([O\,{\sc iii}]/H$\beta~\leq~0.5$) and those near the limit with LINERs;
but this is probably a coincidence, because the predicted ratio
[S\,{\sc ii}]/H$\alpha$ does not match our data 
(see Fig. 1 of Contini et al.\,1998).
A clear correlation between reddening and oxygen abundance in the sample
of Markarian barred galaxies (Consid\`ere et al., in preparation) provides 
further evidence against this possibility which would predict the
opposite correlation.

\section{Consequences of our discovery}

We believe that the low oxygen abundances of SBNGs imply that they
are young galaxies, still in the process of formation.  
This result has deep implications for the
models of galaxy formation and evolution, and may provide a significant
contribution to our understanding of recent {\it Hubble Space Telescope}
and other observations
of galaxies at high ($z > 2$) and intermediate ($z \simeq 0.5$-2)
redshifts. We review below some of the consequences of our result.

\subsection{The hierarchical formation of galaxies}

The theory that is compatible with
all our observations is that of hierarchical formation of galaxies 
by Tinsley \& Larson (1979), modified by Struck-Marcell (1981) 
to include the effect of gas accretion. 
This theory explains the luminosity-metallicity ($L$-$Z$) relation,
and is the only one that predicts a difference in 
oxygen abundance between early- and late-type SBNGs.

It is already well known that ellipticals 
and dwarf spheroidals trace a $L$-$Z$ relation that spans 13 magnitudes in
$M_B$ and a factor of 200 in abundances (Faber 1973; Brodie \& Huchra 1991;
Bender et al.\,1993). Zaritsky et al.\,(1994) have shown that the
spiral and irregular galaxies follow a similar relationship,
which spans over 10 magnitudes in luminosity and a factor of over 100
in metallicity. They noted, however, that the $L$-$Z$ relation
should be steeper for the giant spirals. They could not come up with
a mechanism that could produce it. 

We find that the late-type SBNGs show the same behavior as the giant spirals
in the $L$-$Z$ diagram (Fig. 1), and it therefore seems
to us that the mechanism that explains the difference between the 
early- and late-type SBNGs should also explain the position of
the giant spirals in the $L$-$Z$ diagram. 

\begin{figure}[t]
%\plotone{davoust_fig1.ps}
%\plotfiddle{davoust_fig1.ps}{8truecm}{0}{50}{45}{-150}{70}
\plotfiddle{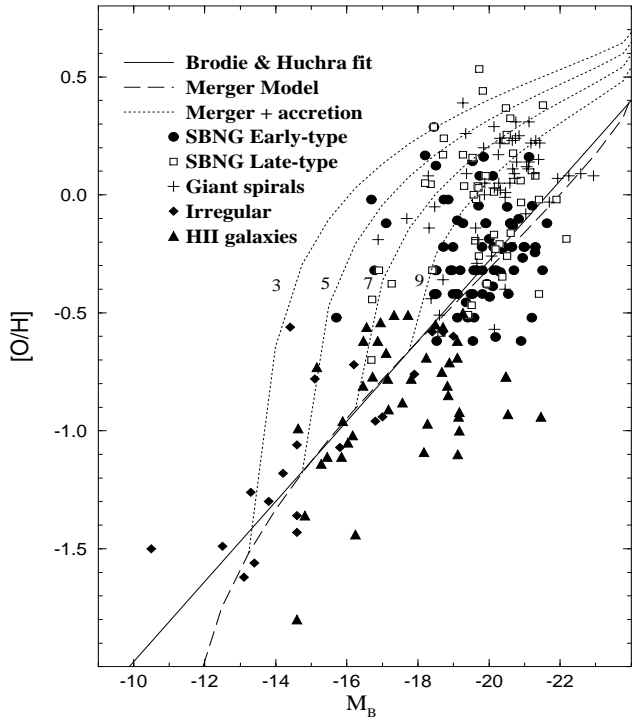}{7truecm}{0}{60}{55}{-190}{-150}
\vspace{1.6cm}
\caption{
The model of hierarchical formation of galaxies applied to the SBNGs.
The solid line is the empirical $L$-$Z$ relation 
for the elliptical and dwarf spheroidal galaxies (Brodie \& Huchra 1991).
The dashed line corresponds to the Tinsley \& Larson (1979) 
merger model.  The dotted lines correspond to
the Struck-Marcell (1981) accretion model after
3, 5, 7 and 9 mergers. A mass-luminosity ratio
of $\sim6$ was used (Faber \& Gallagher 1979).}
\end{figure}

According to Tinsley \& Larson (1979), the spheroidal bulges of spirals 
and S0 galaxies are built in the same way 
as ellipticals, by successive mergers of small masses of gas
and stellar systems, each one producing a burst of star formation.
After each generation of bursts, more stars 
of a given metallicity are created and the metallicities
of the galaxies increase following a $L-Z$ relation.
In a slightly different scenario, 
a galaxy will acquire a disk if, after the violent merger phase producing the
bulge, residual outlying gas or gas-rich small subsystems keep being accreted.
Following Struck-Marcell (1981), accretion of more gas than stars will result 
in a steepening of the mass-luminosity relation, explaining the difference
between early- and late-type galaxies. 
The results of applying Struck-Marcell's models 
to the SBNGs are shown in Fig. 1. 

\subsection{The origin of SBNGs}

The massive nuclear starbursts observed today occur in galaxies that 
merged at an intermediate redshift, i.e. more recently than normal
galaxies. If this is the case,
there should be a population of merging galaxies observable at
intermediate redshifts. These galaxies should be small-mass, irregular,
gas rich, and metal poor systems, probably similar to the nearby HII galaxies. 
This description is consistent with the recent observations 
of the {\it Hubble Space Telescope} which show a substantial increase in
the proportion of galaxies with irregular morphology  and high star 
formation rates at moderately large redshift 
(Madau 1998; Fioc \& Rocca Volmerange 1998; 
Brinchmann et al.\,1998; Lilly et al.\,1998). 

But why can galaxy formation occur so late in some cases?
This is a question for cosmologists, but we can volunteer an explanation:
that the rate of evolution is linked to the local, rather than global, 
density of matter in the Universe.  According to this assumption, 
massive structures such as groups and clusters of galaxies should
be more evolved than field galaxies from the point of view of stellar
populations and metallicity.

\end{document}